\documentstyle[12pt,psfig]{article}

\newcommand{\be}{\begin{equation}}
\newcommand{\ee}{\end{equation}}
\begin{document}
\title{{\bf Thermal photons as a measure for
the rapidity dependence of the temperature}}
\author{
{\bf A.\ Dumitru,
U.\ Katscher,
J.A.\ Maruhn, H.\ St\"ocker, W.\ Greiner}
\\[0.2cm]
{\small Institut f\"ur Theoretische Physik der J.W. Goethe Universit\"at}\\
{\small Postfach 111932, D-60054 Frankfurt a.M., Germany}
\\[0.5cm]
{\bf D.H.\ Rischke}
\\[0.2cm]
{\small Physics Department, Pupin Laboratories, Columbia University}\\
{\small New York, NY 10027, USA}
\\[0.8cm]
{\bf {\small Preprint UFTP 381/1995 \date{March 7, 1995}}}}
\maketitle
\newpage
\begin{abstract}
The rapidity distribution of thermal photons produced in
Pb+Pb collisions at CERN-SPS energies is calculated
within scaling and three-fluid
hydrodynamics. It is shown that these scenarios lead to very different
rapidity
spectra.
A measurement of the rapidity dependence of photon radiation
can give cleaner insight into the
reaction dynamics than pion spectra,
especially into the rapidity dependence of the temperature.
\end{abstract}

One of the goals of heavy-ion physics is the search
for the so-called quark-gluon plasma (QGP), a novel phase of matter where
quarks and gluons are deconfined. It is expected that such a state
can be produced, e.g., when
ordinary hadronic matter is strongly heated or compressed \cite{QGP}.

Real and virtual photons are promising probes of the
QGP \cite{probes} (they may serve, e.g., as a thermometer \cite{KLS})
since they do not suffer from strong interactions. Therefore,
their mean free path is large enough \cite{Thoma} for them
to leave the plasma volume without further interactions.

Transverse momentum distributions of thermal photons produced in $Pb+Pb$
collisions at $160~AGeV$ were already presented in refs.\
\cite{Sriv2,Arb,therm}.
Recently, also the rapidity distribution of electromagnetic radiation was
investigated in more detail \cite{Vogt,Kaempf,Sarkar}. In these works,
however, the collision dynamics was simplified by assuming a
scaling hydrodynamics solution for the longitudinal motion, $v_z=z/t$
\cite{Bj}. This limits the usefulness of the results of
\cite{Vogt,Kaempf,Sarkar} to future collider experiments.
In this letter we present rapidity
distributions of thermal photons in $Pb(160~AGeV)+Pb$ (at vanishing impact
parameter, $b=0$), calculated within the
three-fluid hydrodynamical model, and show that the photons are
very useful to constrain the reaction dynamics at these energies.

Let us first give a brief introduction into the three-fluid
model. For a more detailed presentation, we refer the reader to refs.\
\cite{therm,Uli2,Uli}. The original one-fluid hydrodynamic model \cite{mar85}
assumes instantaneous local thermodynamic equilibrium in the moment when the
nuclei collide and thus is
not appropriate to describe the initial stage of
ultrarelativistic collisions, at least for
$E_{Lab}\ge 10~AGeV$.
This problem is solved here by considering
more than one fluid \cite{ams00}.
The three-fluid model divides the particles involved in a reaction
into three separate fluids: the first two fluids correspond to
the projectile and target nucleons, respectively,
and the particles produced during the reaction are collected in the third
fluid.
Local thermodynamic equilibrium is maintained only in each fluid separately
but not between the fluids. The fluids are able to penetrate
and decelerate
each other during the collision.
Interactions
between the fluids are due to
binary collisions of the particles in the respective fluids.
This allows for a treatment of
non-equilibrium effects in the initial stage of the collision.

The equation of state (EOS) of the target and projectile
fluids is that of an ideal nucleon gas plus
compression energies.
A linear ansatz for the compression energy
with a compressibility of $250~MeV$
and a binding energy of $16~MeV$ is used \cite{EOS1}.

The EOS of the third fluid is that of an ideal gas of massive
$\pi$-, $\rho$-, $\omega$-, and $\eta$-mesons.
At $T_C=160~MeV$ we assume
a first order phase transition into a QGP.
For the (net baryon-free) QGP we use the bag-model
EOS for (pointlike, massless, and noninteracting) $u$ and $d$ quarks.
The bag constant is chosen such that the pressures of both phases
coincide at $T=T_C$.

For comparison, we also perform calculations
within one-dimensional scaling hydrodynamics \cite{Bj} where
we also use the latter EOS. Here, the compressional stage of the
collision is not treated and thus two free parameters,
the initial temperature $T_i$ and
(proper) time $\tau_i$, where the scaling expansion starts, have to be fixed.
We use the values given in ref.\ \cite{KMcLS}: $T_i=300~MeV$,
$\tau_i=0.22~fm$. At later times, the temperature in this model is given by
($T_i\equiv T_i^{(1)}$, $\tau_i\equiv \tau_i^{(1)}$)
\be \label{BjTemp}
T(\tau)=T_i^{(j)}\left( \frac{\tau_i^{(j)}}{\tau}\right)^{c_j^2}\quad,
\ee
where $j=1,2,3$ labels the different phases of matter (QGP, mixed phase,
and hadron gas) and
$c_j$ denotes the sound velocity in the corresponding phase.
All quantities, except for $T_i^{(1)}$ and $\tau_i^{(1)}$,
are determined by the
equation of state, cf., e.g., ref.\ \cite{Sriv2}.

The thermal photon production rate from an
equilibrated, baryon-free QGP is given (to first order in $\alpha$ and
$\alpha_S$) by \cite{KLS}
\be \label{rate}
E\frac{dR^\gamma}{d^3k} = \frac{5\alpha \alpha_S}{18\pi^2}
T^2 e^{-E/T} \ln \left( \frac{2.912E}{g^2 T} +1 \right)  \quad,
\ee
where $E$ is the photon energy in the local rest frame of the fluid.
In the following calculations we fix $\alpha_S=g^2/4\pi=0.4$.
As shown in ref.\ \cite{KLS}, the rate for a
gas consisting of $\pi$-, $\rho$-, $\omega$-, and $\eta$-mesons
may also be parametrized by eq.\ (\ref{rate}). Other contributions, e.g.\ from
the $A_1$ meson \cite{Song}, as well as the effect of hadronic
formfactors \cite{KLS}, are neglected. They are of the
same order of
magnitude as higher order corrections to eq.\ (\ref{rate}), which
have also not been taken into account.
Thus, eq.\ (\ref{rate}) is applied for all phases. The contributions
from the first two fluids are neglected since, for the reactions considered
here, these fluids are cooler. Also,
they undergo a rapid longitudinal expansion and thus
cool faster than the third fluid.
This approximation must, of course, break down at large rapidities, where the
temperature of the third fluid drops rapidly, as discussed below.
As a check, eq.\ (\ref{rate}) has also been applied
to the projectile and target
fluids. This certainly overestimates their contribution because they
contain only baryons. It turns out that, indeed, their contribution to
thermal radiation is negligible up to photon rapidities $\simeq 1.6$.

The thermal photon spectrum
is obtained by an integration over space-time:
\begin{equation} \label{spec}
\frac{d^2N^\gamma}{k_T\,dk_T\,dy}=\int d^4x \, E\frac{dR^\gamma}{d^3k}
\quad.
\end{equation}
Fig.\ 1 shows our results. In the three-fluid model the temperature is
strongly
rapidity dependent and so is the spectrum of thermal photons. As already noted
in ref.\ \cite{therm}, there is much more high transverse-momentum radiation
at midrapidity than in scaling hydrodynamics, which is mainly due to the
different cooling law. In Bjorken's original model, the temperature is
a function of proper time only and
independent of (fluid-) rapidity, and thus the thermal photons show no
rapidity dependence.
This remains true even if transverse expansion \cite{Sriv2,Gersd}
is implemented into scaling hydrodynamics.

\begin{figure}
\vspace*{-2cm}
\centerline{\hbox{\psfig{figure=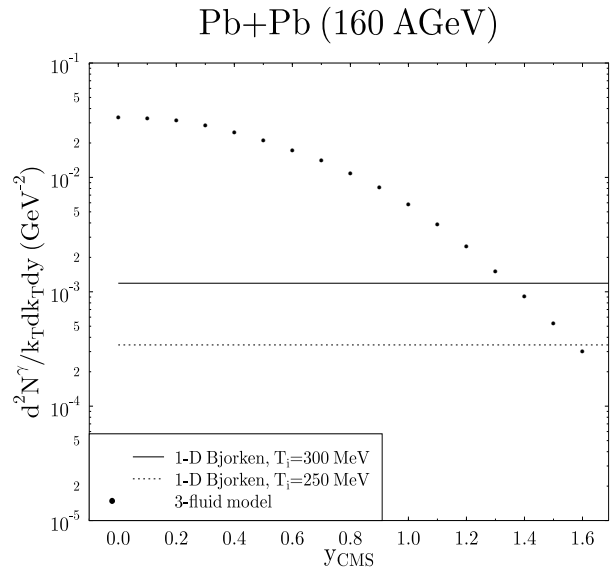,height=11cm,width=15cm}}}

\vspace*{-2cm}
\caption{Rapidity distribution of thermal photons (for $k_T=2~GeV$)
calculated within the
three-fluid model (dots) and longitudinal scaling hydrodynamics.}

\vspace*{.5cm}
\end{figure}

At this point we should comment on the different freeze-out procedures in
the two models. In the three-fluid model the freeze-out of the
third fluid is done instantaneously at some center-of-mass time $t_f$
when its (average) temperature drops below $T_f=100~MeV$.
On the other hand, in the Bjorken model, the freeze-out
takes place at some proper time $\tau_f$ defined by $T(\tau_f)=T_f=100~MeV$.
Thus, the space-time volume for which the photon spectrum (\ref{spec})
is determined, is different in both cases.
However, at least as far as ``hard'' photons are concerned (i.e., $k_T\gg
T_i$) this difference is irrelevant: the shape of the
space-time volume differs only for
the very late stage ($t\approx t_f$), when the temperature is too low to give
a sizeable contribution to the ``hard'' photon yield.

In contrast to the three-fluid model, effects of finite baryon-chemical
potential
(which appears due to the expected large amount of baryon stopping
\cite{Uli2,stop}) on the temperature of the mesons
are not accounted for in the Bjorken model as presented
so far. A finite $\mu_B$ will mainly manifest itself in lower initial
temperatures whereas the {\sl net}\footnote{That is, the change of the
photon rate with $\mu_B$ at fixed $T$.} effect of $\mu_B\neq 0$
on ``hard'' (i.e., $k_T\gg T_i$) photon radiation from the QGP phase
was shown to be small \cite{Du}. We assume that this remains true also in
the hadronic phase. Using the standard argument that relates the final pion
multiplicity to the initial entropy through the assumption
of entropy conservation during the whole expansion stage \cite{KMcLS,Hwa},
and the expression
\be
s=T^3\,\left(\frac{4}{3}\frac{37\pi^2}{30}+\frac{2\mu_B^2}{9T^2}\right)
\ee
for the entropy density in the QGP phase,
it turns out that values $3\le \mu_B/T\le 9$ \cite{Hofm} diminish $T_i$ by
$4-21\%$.
For simplicity let us assume that the ratio of pions to baryons and thus
$\mu_B/T$ is independent of rapidity.
The effect on the thermal photon radiation is
shown in fig.\ 1: the dotted curve was calculated using an initial
temperature of $250~MeV$, all other parameters being the same as before.
If $\mu_B/T$ is taken to be rapidity dependent
this may
introduce some rapidity dependence into the Bjorken model. However,
the resulting photon rapidity spectrum
would still lie between the dotted and the full curve in fig.\ 1.

In principle it would be possible to introduce a (fluid-) rapidity
dependent initial temperature into the Bjorken model by fitting the pion
rapidity distribution {\sl for all rapidities} \cite{Vogt,Kaempf,Sarkar}.
Then, eq.\
(\ref{BjTemp}) is applied in each rapidity slice separately,
i.e.\ $T_i\rightarrow T_i(\eta)$, $T(\tau)
\rightarrow T(\tau,\eta)$, and no longer
globally. We will, however, not adopt this procedure here since it introduces
an infinite number of parameters and contradicts the original philosophy
of scaling hydrodynamics as a simple model with only two free parameters
($T_i$, $\tau_i$). Also, this procedure leads
to a violation of the conservation laws expressed by the hydrodynamic
equations of motion unless a finite (and rapidity dependent) baryon-chemical
potential is introduced at the same time \cite{Kaempf}.
It can be shown that for longitudinal scaling expansion (even including a
cylindrically symmetric transverse expansion)
$\partial p(T,\mu)/\partial\eta=0$, or, equivalently,
\be
s\frac{\partial T}{\partial\eta}+n\frac{\partial\mu}{\partial\eta}=0
\ee
has to hold (the partial derivatives with respect to $\eta$ being
performed at constant $\tau$),
where $s$ and $n$ denote the entropy and particle number, respectively.

\begin{figure}
\vspace*{-2cm}
\centerline{\hbox{\psfig{figure=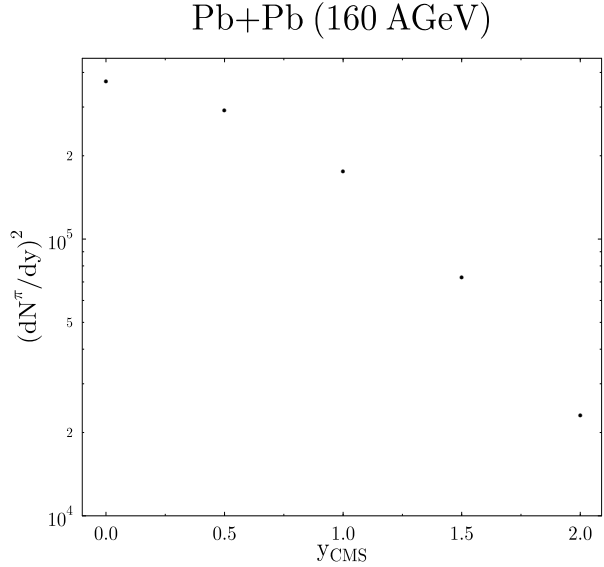,height=11cm,width=15cm}}}

\vspace*{-2cm}
\caption{Squared pion rapidity density.}

\vspace*{.5cm}
\end{figure}

 From simple arguments one may expect the local photon density to
be proportional to the square of the local pion density. However, in the
three-fluid model this proportionality is not maintained in the finally
observed pion spectra. As one can see in
fig.\ 2, the squared pion multiplicity obviously decreases much slower with
rapidity.
Also, the transverse momentum distribution of pions is softer than
that of photons \cite{therm}.
This can be readily explained with the help of fig.\ 3 which shows
the rapidity
distribution of the temperature of the third fluid.
Note that the highest temperatures, which dominate ``hard'' photon and
high-mass dilepton production, prevail only for short times where the
rapidity distribution of produced particles is narrow (in the present model).
This is due to the
fact that the third fluid is initially produced at midrapidity and broadens
in rapidity-space during the expansion.
On the other hand, the final pion distribution emerges at freeze-out where
the temperature rapidity distribution is much broader.
Therefore, the final pion (rapidity) distribution is also much broader
(in rapidity space) than that of the photons.
A measurement of the thermal photon
rapidity spectrum would help to answer if this picture is correct
(at these energies), or if, instead, the temperature distribution is
broad in rapidity space from the very beginning of the expansion (as in
scaling hydrodynamics).

\begin{figure}
\vspace*{-2cm}
\centerline{\hbox{\psfig{figure=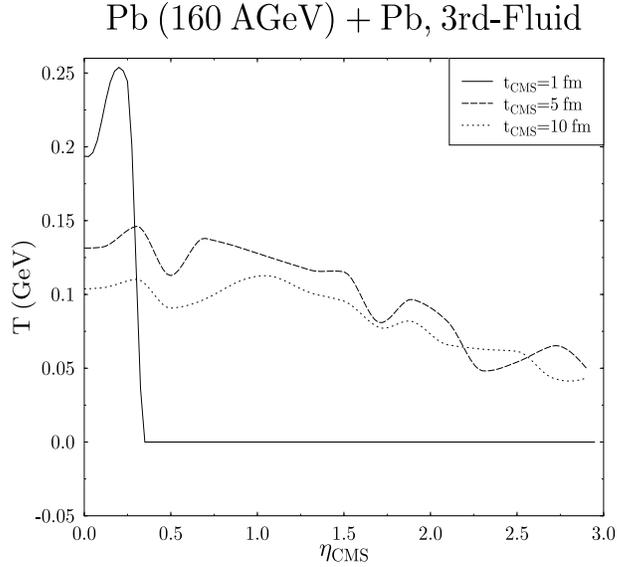,height=11cm,width=15cm}}}

\vspace*{-2cm}
\caption{The (mean) temperature of the third fluid as a function of
fluid-rapidity.}

\vspace*{.5cm}
\end{figure}

In conclusion, the rapidity distribution of thermal photons
in $Pb+Pb$ reactions at the SPS has been studied
within three-fluid and scaling hydrodynamics.
It has been demonstrated that thermal photons provide a powerful tool to
constrain the reaction dynamics, i.e., the time
and fluid-rapidity
dependence of the temperature of the hot and dense reaction zone.

\vspace*{1cm}
{\bf Acknowledgements:}
We
acknowledge helpful discussions with H. Sorge and D.K. Srivastava.
D.H.R. thanks the Alexander v.\ Humboldt-Stiftung for
support under the Feodor-Lynen program and the Director, Office of Energy
Research, Division of Nuclear Physics of the Office of High Energy and
Nuclear Physics of the U.S.\ Dept.\ of Energy, for support under contract
No.\ DE-FG-02-93ER-40764. This work was supported by BMFT, DFG, and GSI.

\end{document}